\documentclass[pdflatex,sn-mathphys-num]{sn-jnl}

\usepackage{graphicx}%
\usepackage{multirow}%
\usepackage{braket}
\usepackage{amsmath,amssymb,amsfonts}%
\usepackage{amsthm}%
\usepackage{mathrsfs}%
\usepackage[title]{appendix}%
\usepackage{xcolor}%
\usepackage{textcomp}%
\usepackage{manyfoot}%
\usepackage{booktabs}%
\usepackage{algorithm}%
\usepackage{algorithmicx}%
\usepackage{algpseudocode}%
\usepackage{listings}%
\usepackage{soul}
\usepackage{cancel} 
\usepackage{enumitem} 

\theoremstyle{thmstyleone}%
\theoremstyle{thmstyletwo}%

\theoremstyle{thmstylethree}%

\begin{document}

\title[Article Title]{Entropy Computing, A Paradigm for Optimization in Open Photonic Systems}

\author*{\fnm{Lac} \sur{Nguyen*}}\email{lnguyen@quantumcomputinginc.com}

\author*{\fnm{Mohammad-Ali} \sur{Miri}*}\email{amiri@quantumcomputinginc.com}

\author{\fnm{R. Joseph} \sur{Rupert}}

\author{\fnm{Wesley} \sur{Dyk}}

\author{\fnm{Sam} \sur{Wu}}

\author{\fnm{Nick} \sur{Vrahoretis}}

\author{\fnm{Irwin} \sur{Huang}}

\author{\fnm{Milan} \sur{Begliarbekov}}

\author{\fnm{Nicholas} \sur{Chancellor}}

\author{\fnm{Uchenna} \sur{Chukwu}}

\author{\fnm{Pranav} \sur{Mahamuni}}

\author{\fnm{Cesar} \sur{Martinez-Delgado}}

\author{\fnm{David} \sur{Haycraft}}

\author{\fnm{Carrie} \sur{Spear}}

\author{\fnm{Joel Russell} \sur{Huffman}}

\author{\fnm{Yong Meng} \sur{Sua}}

\author{\fnm{Yu-Ping} \sur{Huang}}

\affil*{\orgname{Quantum Computing Inc (QCi)}, \orgaddress{\street{5 Marine View Plaza}, \city{Hoboken}, \postcode{07030}, \state{NJ}, \country{USA}}}

\abstract{
Finding better solutions to combinatorial optimization problems could have a large positive impact on many real-world application areas, such as logistics. For this reason, significant efforts have been made to design novel optimisation paradigms. Here we show an early instance of such paradigm in an optical setting, the entropy computing paradigm. Specifically, we experimentally demonstrate the feasibility of entropy computing by building a hybrid photonic-electronic computer that uses optical measurement and feedback to solve non-convex optimization problems. 
The system functions by using temporal photonic modes to create qudits in order to encode  probability amplitudes in the time-frequency degree of freedom of a photon.  This scheme, when coupled with with electronic interconnects, allows us to encode an arbitrary Hamiltonian into the system and solve non-convex continuous variables and combinatorial optimization problems. We show that the proposed entropy computing paradigm can act as a scalable and versatile platform for tackling a large range of NP-hard optimization problems.
}

\maketitle

\section{Introduction}\label{sec1}

In the domain of computational optimization, a significant number of problems are  NP-hard, which is commonly considered the hardest class of optimization problems. It has been shown that there exists a polynomial-time mapping of all problems in NP\cite{cook1971NPcomplete,karp1972reduce} to a smaller set of complete problems, and further to their extension beyond decision problems, to NP-hard problems including optimization. The computer science community largely believes that no polynomial time algorithm exists for all problems in NP (as an efficient algorithm for any of these complete problems, would imply), but this has yet to be proven. Finding high quality approximations (or optimal solutions) in practical timescales is very important, since numerous real world applications requires solutions to NP-hard problems. 
 
The inherently complex nature of such optimization problems necessitates the development of novel computational frameworks, both in terms of hardware and algorithms. Traditional computational methods implemented on Complementary Metal Oxide Semiconductor (CMOS) devices often struggle with the scale and complexity of these problems, leading to extended solution times or, in some cases, the inability to find an optimal solution within a reasonable time frame. 

Various special purpose processors (both classical and quantum), such as analog solvers, have recently been proposed.  A detailed review of some of these machines can be found in \cite{Mohseni2022Ising}. Many of the recently proposed classical special purpose processors are a class of CMOS devices which effectively implement  simulated annealing and related algorithms, a review of these devices can be found here \cite{matsubara2020digital,kao2023digital}. 

Similarly, many quantum devices used for non-convex optimization are quantum annealers, which were first  proposed by Kadowaki and Nishimori \cite{Kadowaki1998annealing}. Quantum annealers differ from the more conventional gate-based quantum computing architectures, since they encode and solve problems via  continuous time evolution. Recent theoretical investigations showed annealing-based architectures might give equivalent advantages to those seen in a gate model setting. Adiabatic quantum computing, an idealized setting of quantum annealing where very long anneal times are employed, was shown to give the same quadratic speedup \cite{roland2002adiabatic} that Grover search yields in a gate-model device \cite{grover1996search,grover1997search}, and is the best that is achievable by any quantum algorithm \cite{bennett1997GroverBest}. Furthermore, a more general class of continuous-time algorithms can obtain the same speedup, including continuous time quantum walks \cite{childs2004spatial} and interpolations between adiabatic and quantum walk protocols \cite{Morley19aqc-qw}. Even more recently, the same advantage has also been shown to be possible through dissipative dynamics, similar to those that our device emulates \cite{Berwald2023Zeno}; similar effects can be used to solve optimization problems \cite{Berwald2023ZenoCompute}. The physical realization of quantum annealers on superconducting platforms remains a challenge due to limitations on connectivity and scalability.

Numerous approaches have been made to make optical analog computers recently due to the advantage of light in energy efficiency, scalability, and global connectivity\cite{Stroev2023AnalogPhotonic}. A popular approach among all is coherent Ising machines (CIM) \cite{Yamamoto2017CoherentIsing, Yamamoto_2020, doi:10.1126/science.aah5178}. Superior time-to-convergence vs. problem size scaling compared to annealers was observed \cite{doi:10.1126/sciadv.aau0823}. However, maintaining stable operation on CIM to avoid external perturbations, including amplitude heterogeneity or phase-stability demand over long fiber, is the main drawback that prevents this technology from being widely adopted \cite{Leleu2019CoherentIsing, PhysRevA.88.063853}.

It is important to note that CIM implementations solely focus on solving Ising-type problems. However, the landscape of computationally-hard optimization problems is not limited to Ising problems and contains, for example, binary, non-convex continuous-variable, integer, and mixed-integer problems. Many discrete NP-problems do not naturally or directly map to the Ising model's framework of binary spin states with quadratic interactions (although they must in a formal sense due to the definition of NP-hardness), and developing an effective mapping that accurately represents the original problem within the Ising framework can be highly non-trivial and is often achieved only with a high computational overhead. Work in this direction includes \cite{Lucas2014Ising} where mappings for many problems were shown, mappings of maximum k-SAT problems for arbitrary $k$ \cite{chancellor2016kSat}, mappings of the weighted maximum independent set problem and related problems \cite{choi2010IsingMIS}. Furthermore, incorporating constraints, which are often required in NP-hard problems, into the Ising model can increase the complexity of the problem formulation. This is typically accomplished by introducing additional spins (qubits in the context of quantum computing) and carefully designing the interaction terms to ensure that the constraints are properly enforced, which can significantly increase the size and complexity of the resulting Ising problem. An additional complication, particularly for analog computers, is that the ratio of the largest to the smallest relevant energy scales (the dynamic range) can be large in some problems, particularly when constraints are considered, or where mapping procedures to an incompatible hardware graph, for example through minor embedding, must be performed \cite{choi08minor}. The small dynamic range of any real hardware often limits the types of problems that can be efficiently solved on that hardware.

In light of these challenges, we present early steps toward a computing paradigm, \textit{entropy computing}, that operates by conditioning a \textit{quantum reservoir} to promote the stabilization of the ground state in an optical setting. In this approach, a target Hamiltonian is mapped onto an effective dissipative operator to solve an optimization problem in the photon number Hilbert space. This approach induces loss into the system to suppress the evolution of unwanted states while promoting the evolution of (qu)dits that represent lower energy states of the corresponding Hamiltonian. Here, we define entropy computing as a system that encodes information in photon number states with readout in the Fock basis, and employs a balance of loss and gain, potentially via measurement and feedback, to search for optimal solutions, following the principle of minimum entropy as discussed in \cite{Vadlamani2020minEntropy}.

We experimentally demonstrate entropy computing through a hybrid optoelectronic measurement-feedback system that utilizes photon qudits encoded as probability amplitudes in time-frequency degree of freedom in conjunction with electronic interconnects for embedding an arbitrary Hamiltonian. In this manner, we demonstrate an entropy computing machine with a versatile polynomial loss function containing first- to fifth-order terms with fully programmable weight tensors, capable of solving optimization problems with up to $949$ independent variables over a fixed summation constraint. We employ this platform for solving non-convex continuous variables and integer combinatorial optimization problems. While the goal of this paper is not to fully introduce the paradigm, we do include some discussion of how a more quantum version which is still within this paradigm could be created in supplementary note 1.

\section{Methods}\label{sec2}

\subsection{Hybrid Entropy Computing System}

\begin{figure}[H]
\centering
\includegraphics[width=0.75\textwidth]{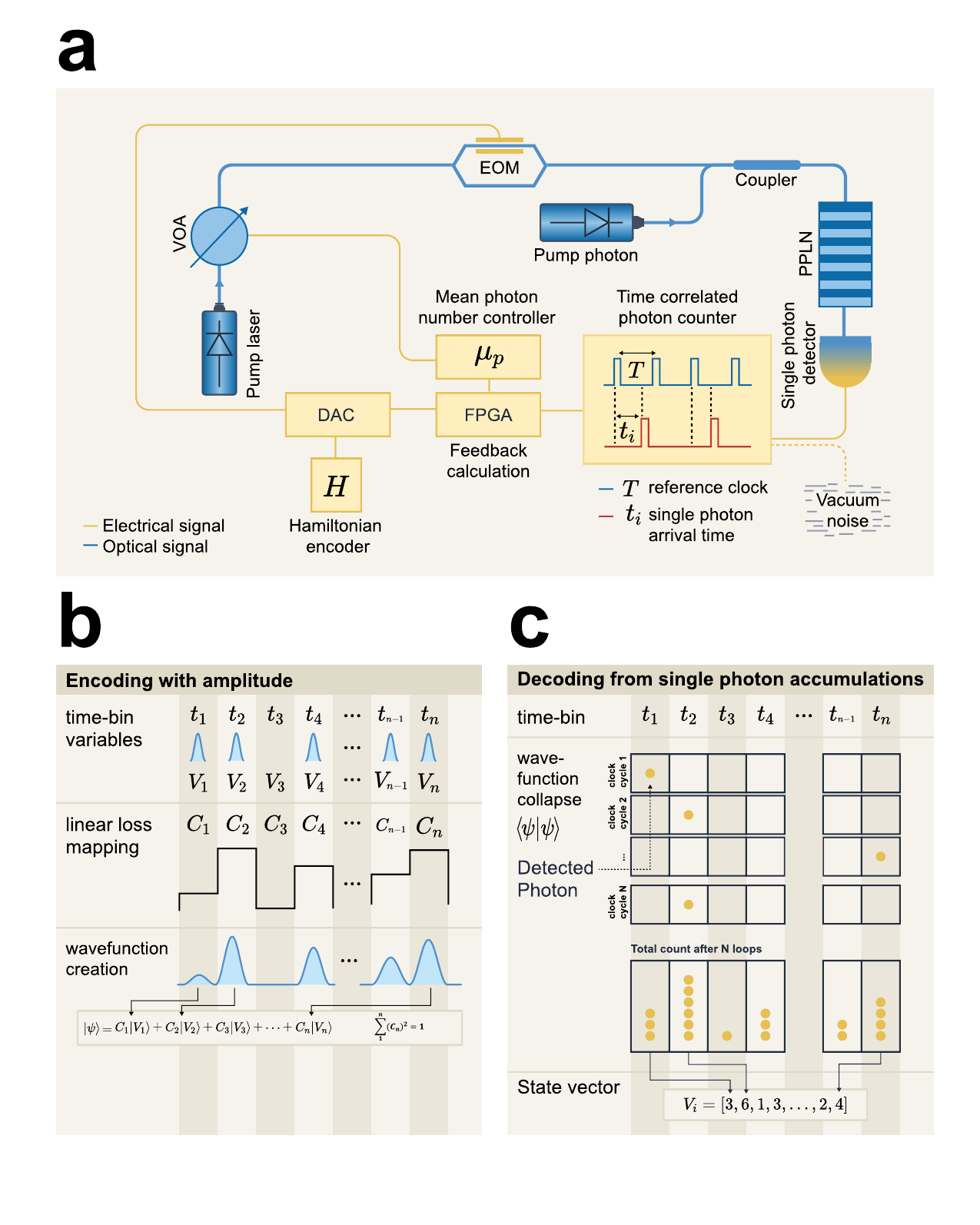}
\caption{\textbf{An emulation system for entropy computing using time-energy modes and a measurement-feedback scheme.} (a) The optical signal is generated by a continuous-wave laser followed by a set of variable optical attenuators (VOA). It passes through the electo-optic modulator (EOM), the periodic-poled lithium niobate (PPLN) for nonlinear process, and is detected by a single photon detector (SPD). The time correlated single photon counting (TCSPC) results are fed to a field-programmable gate array (FPGA) board, where the radio-wave input to the EOM is calculated and generated in a digital-to-analog converter (DAC) for each time bin. Here, the nonlinear circuit (in this case sum frequency generation via PPLN), photon detector, and the FPGA function together as a mixer/encoder. (b) The quantum states are encoded into a train of time-bin states of light in the photon-number Hilbert space. The linear loss in the Hamiltonian is mapped into probability amplitudes of a wave function. Each variable of the objective function is assigned into each time bin, creating ``qudit'' equivalent that is widely used in high-dimensional temporal encoding in quantum random number generation, quantum key distribution, and single-photon sensing \cite{Rehain21, Nguyen18, Mower_2013}. (c) The single photon collapsed states are collected one by one and accumulate to create the next feedback to tailor new wave functions in each loop. When the wave function evolutions reach a stable distribution, the number of photons collected in each bin are translated directly as a state vector multiplied with a normalization factor.} 
\label{fig2}
\end{figure}

In the first realization of this hardware, which is discussed here, we use time correlated single photon counting (TCSPC) and electro-optical feedback to emulate entropy computing, although for discounted computing power. We discuss in supplementary note 1 how an all-optical extension of this paradigm may be realized. The system configuration is illustrated in Fig.~\ref{fig2}, where the loss mechanism is implemented through an electro-optical modulator (EOM) and the mixer is realized by passing the optical signals through a nonlinear optical circuit, detecting them in a single photon detector, and post-processing the TCSPC results. In details, the Hamiltonian problem is encoded into the amplitude of an electrical signal via digital-to-analog converter (DAC). The signal is used to drive an electo-optic modulator (EOM) device which tailors a weak coherent state into single-photon signals in a shaped wave function, to realize high dimensional time-bin encoding. The optical signal is then combined with and modulated by a coherent pump at a different wavelength as they pass through a quantum non-linear optical circuit. In this report, we use a periodic-poled lithium niobate (PPNL) non-linear crystal as sum frequency converter where signal light is single-photon level output from the EOM and pump photons are undepleted to ensure efficient conversion. At the output, the resultant single photons are detected by a single photon detector (SPD) and recorded by a TCSPC. A clock signal is used as reference where the period matches with the feedback loop time. Field-programmable gate array (FPGA) accumulates photon counts of each time-bin and computes the contribution of those counts to the losses of each time bin, thereby emulating the interaction terms in the Hamiltonian. Hence, the loss rate for each mode is the sum of a constant term, corresponding to the linear ``chemical potential energy'' of that mode, and a photon-number dependent term, corresponding to the nonlinear interaction energy. The quantum frequency converter,  time-correlated single photon counter, SPD and feedback through the EOM together act as the medium with linear and nonlinear losses. Single photon detection combined with FPGA normalization feedback promotes least loss states in a hybrid framework. A secondary DAC embedded in the FPGA is used to continuously control variable optical attenuators (VOA), guaranteeing ultra-low mean photon number. 
While our entropy computing approach implementation is designed for photon number states or Fock states, it is practically difficult with the current state-of-the-art in deterministic photon number generation \cite{Uria_2020, PhysRevA.108.053709}. In current hybrid implementation of Dirac-3, we use a common approximation of Fock states which is coherent states with ultra-low mean photon number $\mu$\cite{RevModPhys.74.145}. Dirac-3 is maintained such that about 1$\%$ or less of pulses has more than 1 photon. We further provide evidence that solution quality of Dirac-3 is affected by varying mean photon number. 

In this system, the temporal bins form the state bases and TCSPC measures photon counts in each bin for stochastic computing with integer variables \cite{bouchard2022quantum, kaiser2021probabilistic}. The weak coherent states with low power will approximately yield either vacuum or a superposition of single photons in each time bin, in other words states where two or more photons are present are exceedingly rare.

Figure \ref{fig2}c depicts measurements where one time-bin is occupied after each measurement. Note that while we do often measure single photons, even very weak coherent states do not reproduce single photon statistics, for example they would not be able to reproduce celebrated photon statistics effects such as the Hong-Ou-Mandel effect\cite{Hong1987HOM,Gerry_Knight_2004}. They do however provide us a convenient means to test early hybrid devices acting in the few photon regime  Fluctuations are introduced through the shot noise of single photon counting. Through the feedback, the photon counts of each time bin are used to condition the photon losses applied to the same or other time bins in the subsequent roundtrip iteration. Hence, the quantum states evolve step-wise through a measurement-and-feedback setting. By enforcing a normalization condition for the total photons in all time bins, the photon numbers in each time bin will be incentivized to self-align into a configuration with the minimum loss, as the high loss configurations will be ``punished'' through the loop iteration. As we can conveniently control the loss for individual time bin and the overall loss, one can apply constraints to photon numbers in each time bin or the sum of all. This gives the flexibility of optimizing practical problems that always come with various constraints. 

Figure \ref{fig2}a presents a hybrid entropy computing system that operates on time-bin modes using a measurement-feedback scheme. The optical signal is generated by a C-band continuous-wave laser attenuated using a set of programmable variable optical attenuators (VOAs) to produce weak coherent states. This signal is then modulated by an electro-optic modulator (EOM), driven by a radio-frequency (RF) waveform tailored for each time bin. The modulated, single-photon-level signal is combined with an undepleted O-band pump in a periodically-poled lithium niobate (PPLN) waveguide for a sum-frequency generation (SFG) nonlinear process. The upconverted photons are detected using a high-efficiency, low-dark-count silicon single-photon detector (Si-SPD). Time-correlated single-photon counting (TCSPC) data is sent to a field-programmable gate array (FPGA), which computes the appropriate RF waveform and drives the EOM via a digital-to-analog converter (DAC) on a per-time-bin basis.
In this architecture, the nonlinear circuit (SFG in PPLN), photon detection, and FPGA together form a feedback-based optical mixer and encoder. Figs. \ref{fig2}b and \ref{fig2}c show that, over multiple feedback loops, the accumulated photon counts in each time bin represent the values of corresponding variables. The resulting time-bin histograms at each iteration reflect the state vector of the Hamiltonian at that moment.
 For accurate operation, it is critical to maintain an ultra-low dark count rate in the SPD so that the measurement is dominated by shot noise. The speed of single-photon detection is a key factor influencing the overall feedback latency. To ensure a compact and stable system, PPLN is used to leverage the high efficiency and low dark count performance of the Si-SPD.

Coherent states quantum state of light closely resemble classical oscillation of light field, but also include  quantum noise, specifically shot noise. In this hybrid system, single photon count of each time-bin experience Poisson noise which is taken advantage as a source of ``entropy'' or ``fluctuation'', in each feedback loop. We define this parameter as ``quantum fluctuation coefficient'' that equals to $\frac{1}{\sqrt{N}}$, following the standard deviation of Poisson distribution where $N$ the number of photon collected in each time-bin. Therefore, throughout the measurement and feedback process, the injection of  fluctuations enables a search of the solution space, which assists in bypassing the trapping of the system into local minima. The results section of this paper evidence solution quality of Dirac-3 affects by varying quantum fluctuation or shot noise. We expect that an all optical system could perform even better and we discuss a route to one in supplementary note 1. It requires many incremental loops to find the global minimum solutions to optimization problems with many variables. It is worth noting that since the optical parts of the system do not involve large scale non-Gaussian quantum superpositions, they could in principle be instead classically simulated, as has been argued in the coherent Ising machine setting \cite{Clements2017GaussIsing,Tiunov2019SimCIM}. In spite of this fact, these machines still provide a useful demonstration of what can be done in optical computing, and a preview of what a fully optical system may be able to do. The ultimate goal of a quantum version of an entropy computer would be to take full advantage of the non-Gaussian statistics induced by Fock-state measurement, this is discussed in supplementary note 4.

\subsection{Dirac-3 Optimization Machine}
The hybrid system can be used for a variety of optimization tasks, including those of binary variables, mixed-integer variables, quasi-continuous, and any combination of them. In this work, we focus on the minimization of the following cost function $E$ over variables $v_i$:
\begin{align}
    E = \sum_{i} C_i v_i +
        \sum_{i,j} J_{ij} v_i v_j +
        \sum_{i,j,k} T_{ijk} v_i v_j v_k + \nonumber \\
        \sum_{i,j,k,l} Q_{ijkl} v_i v_j v_k v_l +
        \sum_{i,j,k,l,m} P_{ijklm} v_i v_j v_k v_l v_m ~ .
\label{eqE}
\end{align}
Here, $v_i$ ($i=1,2,3,\cdots, N$) are real numbers over a discrete space, 
$C_i$ are the linear terms' real-valued coefficients while $J_{ij}$, $T_{ijk}$, $Q_{ijkl}$, $P_{ijklm}$ represent two-body to fifth-body interaction coefficients that are real numbers subject to the tensors $J$, $T$, $Q$, and $P$ being symmetric under all permutations of the indices. Furthermore, it is important to note that the probabilistic nature of the optimization variables in the proposed entropy computing method demands that all variables are non-negative, i.e., $v_i \geq 0$. To allow negative variables, a linear transformation of the variables needs to be applied prior to the problem encoding. While we are aware of proposals to implement higher order interactions in similar optical settings (for example \cite{Khairul2023HighOrder}), we are not aware of any other experimental implementations reported in the literature. It is also worth noting that we do not necessarily need dense higher order interactions to encode interesting problems. As a concrete example here the hardest satisfiability problems occur when the number of clauses scale with the number of variables \cite{Selman1996hardSAT}. This concept seems to hold when quantum heuristics are considered as well \cite{Santra2014Annealing2SAT}.

In observation of the system openness due to effective dissipation and gain applied during each feedback loop, a constraint on the sum of the variables $v_i$ can be applied such that: 
\begin{equation}
    \sum_{i} v_i = R.
\label{eqR}
\end{equation}
Thus, the proposed scheme finds minimal solutions of the loss function of Eq.~\eqref{eqE} subject to the sum constraint of Eq.~\eqref{eqR}. 

Given a desired optimization problem formulated in the form of relation~\eqref{eqE}, the optimization process is described below. First, a sum constraint $R$ is chosen. Clearly, without a knowledge of the final solution, the true sum of variables $\sum_i v_i$ is unknown. But, one can readily bypass this problem by introducing slack variables, as discussed in an example in the next section. Once the sum is predetermined, an initial photon qudit state is generated from the noise and launched in the system. After propagating in the closed loop, the photonic state is measured and evaluated to prepare the state for the next iteration.\\

In stark contrast with the Ising Hamiltonian, which is the basis model for the majority of quantum annealers, the above objective function involves polynomial terms (up to fifth order in our present experimental demonstration) over discretized variables. In this regard, the proposed hybrid quantum optimization machine offers two immediate advantages over an Ising solver; (i) it can naturally represent higher than binary optimization problems, and (ii) it involves k-body interaction terms ($k=1, 2, 3, 4, 5$). 
Accordingly, it offers great potential in efficiently solving continuous and integer variables as well as problems that naturally involve higher-order interaction terms such as the satisfiability boolean, without requiring additional complex encoding or incorporating auxiliary variables that add to the size of the problem in case of an Ising solver. Furthermore, the proposed machine naturally allows for dense long-range interactions in all orders of the interactions, which alleviates the requirement for complex embedding algorithms. Here, we report results from our first commercially available machine, which we call Dirac-3. Dirac-3 is an optimization solver which implements the entropy computing paradigm discussed above.

\section{Results \label{sec:expt_results}}

In the following, we first use a simple 2-variable problem to illustrate the optimization process in  our early hybrid implementation of our entropy computing paradigm. Next, we present results from two benchmark hard optimization problems from known libraries. In general, benchmarking systems like ours is a challenging problem. We do however present some promising experiments to show how the technology works. Here, we use two different types of problems, a non-convex quadratic problem, and a combinatorial graph cut problem.

\begin{figure}[h]
\centering
\includegraphics[width=1\textwidth]{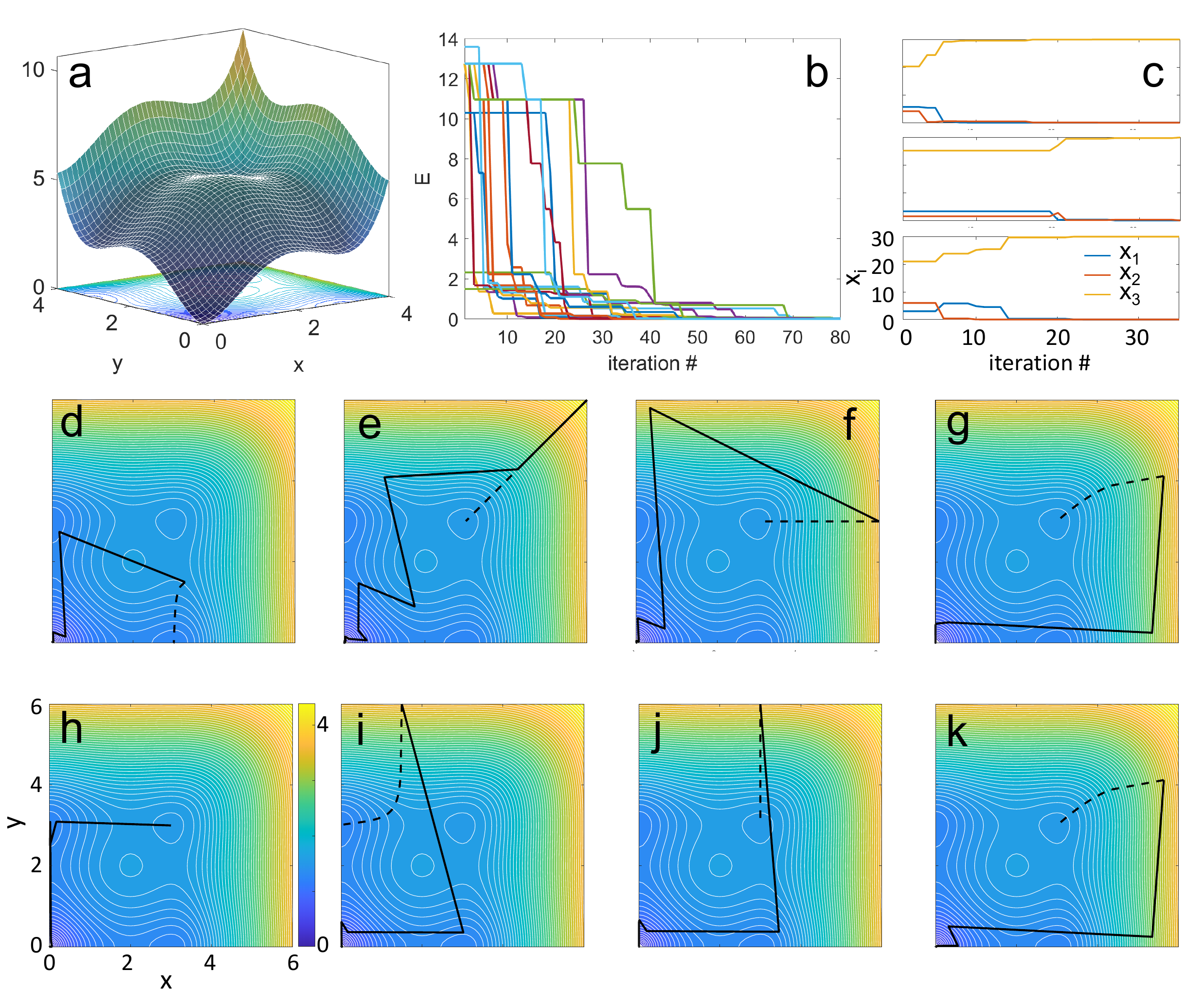}
\caption{\textbf{Solving a two-variable non-convex quadratic optimization problem.} A two-variable non-convex polynomial optimization problem is considered. (a) A visualization of the energy landscape that involves three local minima and a global minimum at $(\textup{x},\textup{y})=(0,0)$. (b) The iterative evolution of the cost function of the proposed hybrid entropy computing solver over 20 runs. (c) Three exemplary evolutions of the optimization variables involved, including the slack variable ($x_3$), over iterations of the entropy computing solver. (d-k) Eight exemplary trajectories of the optimization variables in the two-dimensional $(x,y)$ plane as the solver evolves toward equilibrium while starting from different initial conditions. In these figures, the solid lines show the trajectory of the entropy-computing solver, while the dashed lines depict the trajectory of a gradient-descent solver. More details on the gradient descent method can be found in supplementary note 3.}
\label{fig3}
\end{figure}

We first consider a simple two-variable optimization problem, defined through the objective function $g(x,y) = (x^4+y^4)/4 - 5(x^3+y^3)/3 + 3 (x^2+y^2)$. As depicted in Fig.~\ref{fig3}(a), this function contains three local minima at $(\textup{x},\textup{y})=\{(0,3), (3,0), (3,3)\}$ as well as a global minimum at $(\textup{x},\textup{y})=(0,0)$. To solve this problem using Dirac-3, we consider 3 variables $v_1$, $v_2$, and $v_3$, with $v_1$ and $v_2$ representing $x$ and $y$ respectively, while $v_3$ representing a slack variable which is not coupled to any other variable but used to effectively relax the sum constraint. In this manner, one can consider a relatively large sum constraint of $R=100$ to incorporate a larger range of potential solutions. Figure~\ref{fig3}(b) shows the evolution of the energy level for $20$ different runs as the system evolves toward an equilibrium point. The evolution of the variables versus the number of iterations are shown in Figs.~\ref{fig3}(c) for three exemplary cases. These figures indicate rapid convergence of the solution after only a few iterations. Even more interestingly, the trajectories of the optimization variables in the two-dimensional $(x,y)$ plane (Figs.~\ref{fig3}(d-k)) show that the system takes steps to avoid the local minima, even though in all cases the initial point was located within the attractor basins of the local minima.

\begin{figure}[h]
\centering

\includegraphics[width=0.9\textwidth]{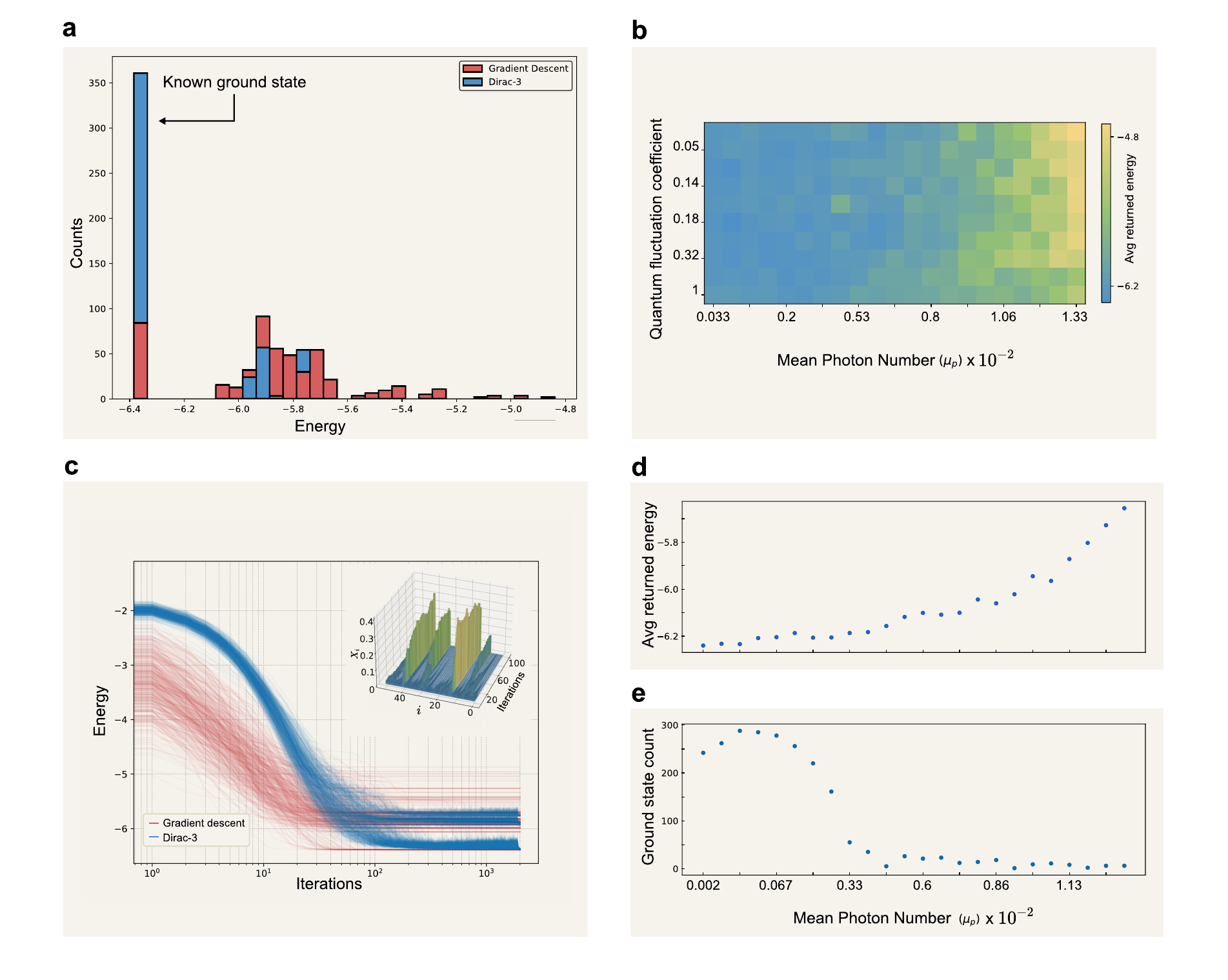}
\caption{\textbf{Solving a non-convex continuous optimization problem.} A non-convex quadratic optimization problem with 50 variables is considered (QPLIB$\_$0018). (a) Energy distribution over 500 runs of Dirac-3 (blue) and gradient descent algorithm (red). (c) Energy evolution versus the number of iterations on Dirac-3 (blue) and gradient descent (red). The inset shows the evolution of the solution versus iterations. (b) Relationship between the mean photon number ($\mu$), quantum fluctuation coefficient,and key performance metrics - average returned energy. The transitioning from left to right, represents an increase in mean photon number ($\mu$), showing the probability that Dirac-3 is operating in single-photon regime. Quantum fluctuation or shot noise is used as $\frac{1}{\sqrt{N}}$ where N is the number of photon count accumulated in each time-bin. From bottom to top of the vertical axis, this coefficient is gradually decreased. (d), (e) Average returned energy and number of ground states found after 500 runs are collected when mean photon number is decreased further up to $0.002\%$, close to the level of dark count of single photon detector.}
\label{fig4}
\end{figure}

We next consider a non-convex quadratic optimization problem (QPLIB$\_$0018) with $50$ continuous variables over a fully connected weighted graph, selected from QPLIB, a library of quadratic programming instances \cite{furini2019qplib}. The cost function of this problem is of the form $f(\mathbf{x})=C^T \mathbf{x} + \mathbf{x}^T J \textbf{x}$, where, $\textbf{x}=(x_1, x_2, \cdots, x_{50})^T$ and $x_i \geq 0$. Figures~\ref{fig4}(a) show the equilibrium energy distribution over 500 runs of Dirac-3 compared with the results obtained from a conventional gradient descent algorithm. Dirac-3 successfully finds the ground state in almost $80\%$ of instances. The evolution of the energy level is plotted as a function of the iterations (red) and compared with those obtained from gradient descent (blue) over 50 different runs, shown in Fig.~\ref{fig4}(c). The inset depicts the evolution of the solution with the number of iterations indicating rapid convergence of the solution after a few iterations. An important observation here is that the problem is hard enough for gradient descent to become stuck in a sub-optimal solution, but that Dirac is able to avoid these local minima and find the true optima. This shows that the device is able to do more than just perform updates based on local information from flipping single bits, an encouraging sign for the method.\\

It is important to evaluate the performance of the entropy computing machine while operating with classical versus quantum states of light. We currently use (weak) coherent states as input, while weak coherent states by themselves always behave classically regardless of average photon number, interactions with nonlinear elements can yield quantum states within the device itself, therefore discussion of quantum states is justified. However, very strong nonlinearity (stronger than is currently available) would be needed to be able to achieve all optical implementation, see discussion in supplementary note 1. For this purpose, we investigate the optimization performance for various mean photon numbers $\mu_p$ associated with both regimes. As the same time, we explore how solution quality is affected when shot noise is varied in each feedback loop. It is expected that high probability of operating in the single-photon regime and high shot noise is required in every feedback loop to constantly boost the stochasticity of the system, promoting chances of jumping out of local minima.

We perform the experiment using the same non-convex quadratic optimization problem (QPLIB$\_$0018) from the library for programming instances. In the current Dirac-3, we varied $\mu_p$ from $0.03\%$ to 1.33$\%$ which is equivalent to 98.67 $\%$ to 99.97 $\%$ probability of single-photon regime. Note that similar strategies, but not at such extremely low photon numbers have shown promise in coherent Ising machines \cite{Kumagai2025singlePhotonCIM}. Figure~\ref{fig4}(b) shows the average energy level returns after $20$ runs versus the various mean photon number and quantum fluctuation coefficient. Each pixel represents the average energy returns (Fig. \ref{fig4}b) after 20 runs. Thus, this experiment provides the evidence that better solutions or lower average energy returns for an optimal range associated with lower mean photon numbers. By increasing $\mu_p$ the solution quality deteriorates as the system is more likely to be trapped in local minima. On the other hand, lower $\mu_p$ promotes more single photon states, leading to higher probabilities of escaping local minima and converging to the global minimum. Furthermore, it is important to note that there is a trade-off between increasing the quality of the solution by operating in the low-photon-number regime and reducing the optimization time. Operating at lower photon numbers requires a longer time to accumulate enough photons, leading to an increase in time for the system to stabilize in an equilibrium state. 
Notice that further reducing the single photon rate the classical (thermal) noise from the single-photon detector's dark count deteriorates the results, too. This can be quantified into a parameter, the quantum-to-classical ratio (QCR), defined as the photon detection rate divided by the dark count rate, which decreases for lower photon numbers. Fig.~\ref{fig4}(d),(e) captures the number of ground states found after $500$ runs when decreasing $\mu_p$ ( from right to left) when optimizing QPLIB$\_18$ problem, ground states are found more often. However, when the average photon count reaches the dark count level of the SDP, number of ground states found reduces accordingly. 

Solution quality evidently changes while studied with various quantum fluctuation coefficient. Values of the vertical axis on Fig. \ref{fig4}(d) shows the percentage of shot noise presents when collecting photon in each time-bin. For example, $0.1$ quantum coefficient fluctuation means $10\%$ of the photon count measured in across all time-bin comes from Poisson noise. In this experiment, increasing shot noise (top to bottom) shows improvement in average returned energy over multiple runs. This also explains  the evolution trajectory of Dirac-3 in searching for optimal solution in a non-convex energy landscape shows ``quantum-tunneling-like'' behavior in Fig. ~\ref{fig3}. As the system appears to be able to escape local minima highly effectively. Notice that true tunneling would require a large scale quantum superposition which is not realizable in a hybrid setting. However, numerous classical systems can illustrate behavior which is similar to tunneling and matches tunneling rates. This can be seen in \cite{Muthukrishnan2016Cascade} through the physics of a semi-classical diabatic cascade, or in the ability of path-integral quantum Monte Carlo (a classical algorithm) to match the tunneling rates through a simple barrier \cite{Isakov2016QMC}. 

 In this hybrid optimization machine implementation, FPGA can be conveniently engineered to accommodate these higher order of interaction terms, making this architecture practical with current technology \cite{Dou200564bitFF, NOBLE202484}. Thus, on this hybrid measurement-feedback based approach, time-to-convergence depends partially on the FPGA and its architect for multiplier. It is worth emphasizing that this current FPGA-feedback based architecture cannot support large scale quantum superpositions, similar to measurement-and-feedback based coherent Ising machines \cite{Motta2020CIMLOCC}. However, such an early implementation can provide an test of how the algorithms can perform even in the absence of such correlations, being able to perform well in this setting is valuable evidence for how well an eventual device based on optical interference could preform. While the use of the FPGA is a key limitation, we believe that relatively early (in the implementation of the paradigm) tests such as performed here provide an important indication as to whether the direction is viable to support highly performant optimization, in this case the results are suggestive that it indeed can.

Next, we consider a combinatorial optimization problem, the maximum cut (max-cut) problem, that is known to be NP-hard. Considering for simplicity an unweighted graph, the max-cut is a partition of its vertices into two complementary sets, such that the number of edges between the two sets is maximal. This problem can be generalized to maximum k-cut (max-k-cut), where, the vertices are partitioned into $k$ disjoint subsets, such that the number of edges between the $k$ subsets is maximized. Although the proposed entropy computing paradigm deals with continuous-valued variables in general, it can be utilized to solve such discrete problems by suitable encoding of a discrete (e.g., spin) degree of freedom in the continuum, which can be done in different ways. One way is to map a binary variable $s_i \in \{0,1\}$ onto a vector $\vec{\mathbf{s}}_i = x_i \bigl(\begin{smallmatrix} 1\\0 \end{smallmatrix}\bigr) + y_i \bigl(\begin{smallmatrix} 0\\1 \end{smallmatrix}\bigr) $. In this manner, one can embed a classical bit onto the continuum and identify the two states by taking the maximum of $x_i$ and $y_i$ variables. Furthermore, the values of $x_i$ and $y_i$ can be regularized by considering a regularization term that can be chosen as $x_i+y_i=1$ in its simplest way. This approach can be readily extended to consider a ternary digit (trit), a quaternary digit (quit), and so on, which allow for tackling a general $k$-state standard Potts problem with the proposed entropy computing machine.

Now, considering a graph with $N$ nodes, described with adjacency matrix $A$, i.e., $A_{ij}=1$ for adjacent nodes and $A_{ij}=0$ otherwise, the max-k-cut problem can be formulated as minimizing the following objective function: $E = \sum_{i,j} A_{ij} \vec{\mathbf{s}}_i \cdot \vec{\mathbf{s}}_j + \lambda \sum_{i} {\left \| \vec{\mathbf{s}}_i - \mathbf{1} \right \|}_{1}^{2}$, where the second term is an L1-norm regularization term with a parameter $\lambda$. It is straightforward to cast this objective function in the form of relation~\ref{eqE}. This results in a quadratic function involving $qN$ variable, where $q$ is the dimension of a single artificial spin state and $N$ is the size of the graph.

\begin{figure}[h]
\centering
\includegraphics[width=1\textwidth]{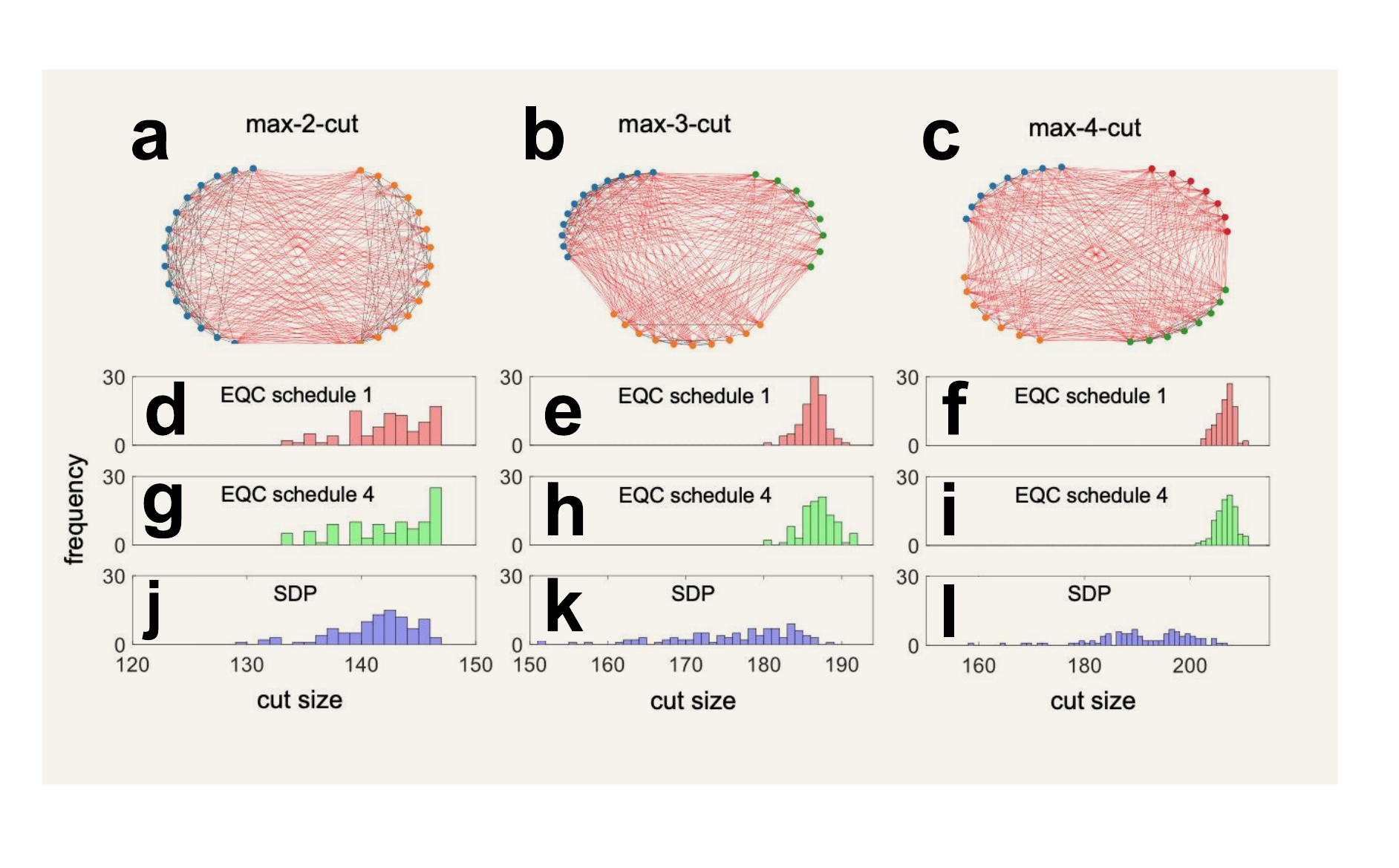}
\caption{\textbf{The results for solving the max-k-cut problem.} The optimization results for solving combinatorial problems of max-cut, max-3-cut, and max-4-cut on a 30-node unweighted graph that is generated randomly with $p=0.5$ probability of connectivity between each two nodes. (a-c) The visualizations of exemplary graph cuts with different numbers of partitions. (d-f) The distribution of the cut sizes over 100 runs of the max-2-cut (traditional max-cut), max-3-cut, and max-4-cut problems on Dirac-3 using relaxation schedule 1. (g-i) Similar distributions when using the schedule 4. (j-l) The distribution of the cut sizes over 100 runs of the same problems using semi-definite programming (SDP). 
}
\label{fig5}
\end{figure}

To test this formulation, we consider a randomly generated graph of $N=30$ nodes such that each two nodes are connected with a probability of $p=0.5$. Figure~\ref{fig5}(a-c) visualizes exemplary partitioning of this graph into $2$, $3$ and $4$ groups through the proposed entropy computing system. This particular graph has $233$ total links, and we found the maximum cut size to be $146$. Histograms of the cut sizes obtained for 100 different runs of the max-cut, max-3-cut, and max-4-cut problems using the entropy computing solver at schedule 1 are shown in Figs.~\ref{fig5}(d-f), and using schedule 4 in Figs.~\ref{fig5}(g-i). The results are compared with those obtained from Semi-Definite Programming (SDP) shown in Figs.~\ref{fig5}(j-l). Here, we used the CVXPY package with the SCS solver to solve the semidefinite program. As expected, we observe that schedule 4 provides better results. In the version of Dirac-3 reported here, schedule 1 was set to have smaller number of feedback loops, lower quantum fluctuation coefficient, and higher mean photon number compare to schedule 4. In addition, as this figure clearly indicates, in all cases the entropy computing machine provides excellent results that outperform SDP results. As a reference, it is worth mentioning that SDP relaxation for max-cut has a performance guarantee of the ratio of $0.878$ \cite{goemans1995improved}, which translates to a cut size of $128$ for the particular graph considered here. For the results presented in Fig.~\ref{fig5}, the regularization parameter was chosen to be $\lambda=5$.

\section{Discussion}\label{sec3}

We experimentally demonstrated entropy computing through a hybrid photonic-electronic system that builds on time-multiplexed photon qudit time bins propagating in a closed feedback loop involving electronic interconnects for implementing a reconfigurable effective optimization cost function. We demonstrated the successful operation of the proposed system for solving non-convex and combinatorial optimization problems. Results show that Dirac-3 found the ground state more often than classical gradient descent on a non-convex optimization problem with constraint. Dirac-3 also performs superior to semi-definite programming in solution quality on Ising and standard Potts problems. The presence of the sum constraint in the current Dirac-3 machine becomes an advantage in problems that intrinsically involve this natural restriction. Such problems emerge in portfolio optimization, resource allocation problems, diet problems, knapsack, network flow problems, and election and voting systems \cite{9810536, ji2024algorithmoriented, witt2024ilpbased, 10.3389/fict.2017.00029}. One of the key strengths of the proposed entropy computing machine is its flexibility in encoding, allowing it to handle continuous and integer variables. This capability sets it apart from many classical and quantum solvers that are primarily designed for binary Ising/QUBO problems, and it opens up new possibilities for efficient solutions in diverse applications, including grid optimization and machine learning tasks like clustering and decomposition \cite{hua2022highorder, Kumar_2018}. Traditionally, implementing higher-order interactions in Ising, Potts, or XY problems on analog hardware solvers requires a quadratization step for polynomial reduction \cite{chancellor2016kSat,Chancellor2019DomainWall,chang2020integer}. Dirac-3 simplifies this process by directly mapping high-order optimization problems, promising increase in precision and solution quality while consuming fewer resources by eliminating the need for auxiliary variables \cite{Bybee2023, Sharma2023}. Further studies and benchmarking are necessary to fully explore the capabilities of Dirac-3.

The speed of our approach comes from relaxation mechanism as well as the fast propagation and processing of information within light. Our current Dirac-3 hybrid implementation does not utilize quantum entanglement. Rather, we demonstrated that our approach to analog optimization is capable of escape local minima in dense NP-hard problem and perform well, using intrinsic randomness of quantum fluctuations in photon number, an early demonstration key aspect of our paradigm of entropy computing. Recently, some studies have demonstrated that taking advantage of quantum superposition, which extends from the double-slit experiment, key role to discovery of quantum mechanics, provides advantage in computation, machine learning, and imaging tasks \cite{Li2024FirstPM,Taddei2021,Zhang:20}.

The current hybrid architecture of the entropy computing system exhibits low energy consumption (below $100$ W during operation), comparable to a laptop. Further benchmarking needs to be done to appreciate energy advantages of this architecture. This energy footprint is expected to decrease significantly when the system is implemented entirely on an integrated photonic platform. This shift towards an all-optical approach holds promise for a practical and sustainable unconventional computing paradigm, contributing to solving energy rising issues by high performance computing \cite{electricity2024, Andrae2015}.

\backmatter



\bmhead{Acknowledgements}

The authors would like to thank other scientists at Quantum Computing Inc including Prajnesh Kumar, Jeevanandha Ramanathan, and Malvika Garikapati for their valuable engineering suggestions. NC was fully supported by QCi in performing this work, in particular no UKRI support was received for writing this paper. The authors thank Mark Campanelli for providing software support for the execution of these experiments.
\textbf{Competing Interests.} All authors hold shares or options in QCi. Y.~Huang is named on a related patent: ``Super Ising emulator with multi-body interactions and all-to-all connections'' which has numbers US20230185160A1 (USA), WO2021231794A1 (worldwide), JP2022569453A (Japan), EP4150430A1 (Europe), KR20230011342A (South Korea), CA3178484A1 (Canada), CN115698897A (China). Y.~Huang is in the process of applying for an additional process related to underlying technology.

\section*{Declarations}

\textbf{Code Availability.} 
Code associated with this paper can be found at \url{https://github.com/qci-github/eqc-studies/blob/main/eqc-paradigm/eqc_vs_grad/grad_vs_eqc.py}. The same code is available on a figshare repository at \url{http://doi.org/10.6084/m9.figshare.29949722}.

\noindent
\textbf{Data Availability.} 
Data associated with this paper can be found at \url{https://github.com/qci-github/eqc-studies/blob/main/eqc-paradigm/eqc_vs_grad/grad_vs_eqc.py}. The same data are available on a figshare repository at \url{http://doi.org/10.6084/m9.figshare.29949722}.

\noindent
\textbf{Author Contribution.} Y.~Huang conceived the concept of the hybrid entropy computing hardware. L.~Nguyen designed and built the computing hardware. Y.~M.~Sua, R.~J.~Rupert, S.~Wu, N.~Vrahoretis, P.~Mahamuni, C.~Martinez-Delgado, D.~Haycraft, and C.~Spear contributed to the development, construction, and testing of the computing hardware. D.~Haycraft and L.~Nguyen designed the experiments to characterize the system and interpreted the results. M.-A.~Miri, W.~Dyk, and I.~Huang conceived and designed the problems and algorithms used to study the computing hardware.  M.-A.~Miri, N.~Chancellor, L.~Nguyen, Y.~Huang, and M.~Begliarbekov  drafted the manuscript. R.~Huffman contributed to visualization by designing and producing images, diagrams, and graphical representations of the concepts and results. N.~Chancellor, M.-A.~Miri, and U.~Chukwu revised, organized, and finalized the manuscript. 

\noindent

\bibliography{sn-bibliography}

\part*{Supplementary Information}

\maketitle

\section*{Supplementary Note 1: All-optical Entropy Computing System}

The working principle of an all-optical extension of the system reported in the manuscript is schematically depicted in supplementary figure \ref{fig1}(a). It consists of an optical-feedback loop that includes an optical amplifier, a photon-mode mixer/encoder, and a loss medium to implement both linear and nonlinear loss mechanisms. Quantum states are encoded as photon numbers in a train of time-bin optical pulses in the loop, which could be either in fiber or free space. During each loop, optical signals are amplified and sent into the mixer that performs linear transformation of the time-bin modes according to the desirable Hamiltonian $\hat{H}$. An example of the mixer uses a series of beamsplitters, optical delay lines, and optical switches, all controlled opto-electronically to construct a multiport, delay-switched Mach-Zehnder interferometer. The mixed signal is then sent to the loss medium, so that the optical system realizes the imaginary time evolution of an open quantum system with linear and nonlinear losses. This loss medium can be implemented via Quantum Zeno blockade where the loss must be stronger than the cavity coupling\cite{huang2012antibunched, McCusker_2013, Huang2010}, or sum frequency generation, promoting dissipation in which the system would gradually relax to the effective ground state. The eventual goal is a device which can leverage quantum effects a ``quantum'' entropy computer.

 In addition to the key aspects we already discussed, a quantum entropy computer would additionally have:

\begin{enumerate}
\item Encoding of the optimization problem in an all optical way, for example through interference in a Mach-Zehnder mesh
\item Controllable loss and gain performed in a way which preserves quantum coherence for example through non-linear optical techniques. Note that this precludes a direct measurement-and-feedback approach because such an approach would involve direct measurement of the quantum information being processed.
\end{enumerate}
Note that a combination readout in the Fock basis  as discussed in the main text and (2) from here leads to a fundamentally non-Gaussian system (analogous to Gaussian Boson sampling). This avoids a fundamental weakness of the CIM paradigm, where the non-Gaussianity arises incidentally rather than as an intentional element in the design. With these clarifications added, we feel that we are justified in discussing entropy computing and by extension quantum entropy computing as paradigms as we have defined them more clearly.

It is worth discussing the similarities and differences between our method and gate model algorithms based on imaginary time evolution, as discussed in \cite{Jones2019Spectra,Motta2020Imaginary,Kosugi2022Imaginary}. The key difference in our paradigm is that it is a special purpose analog machine to perform computation in a method similar to imaginary time evolution, in contrast to compiling a simulation of imaginary time evolution to a universal system. In the near term, special purpose analog machines show great promise, as the engineering requirements are less restrictive than a fully universal machine. 

An example realization of quantum entropy computing using time-energy modes and its schematic in a fiber-optical loop is depicted in supplementary figure \ref{fig1}(a),(b), and (c). The working principle is based on carving a desired Hamiltonian ($\hat{H}$) into a dissipation operator ($-\mathrm{i}\hat{H}$) which tends to favorably support the evolution of a quantum state of light that resembles the lower-energy states, particularly the ground state, of the target Hamiltonian, while simultaneously undermining other states through decay/decoherence. The time-domain evolution can be discretized in a fiber-optical loop setting to iteratively evolve the state and relax the system. The quantum states are encoded into a train of time-bin states of light in the photon-number Hilbert space. During each loop, optical signals are coupled with vacuum fluctuation and amplified by an optical amplifier with fixed average power. The output is sent into the mixer and encoder that performs linear transformation of the time-bin modes according to the desirable Hamiltonian. The mixed signals are sent to the loss medium which causes differential loss to each time mode before subject to the constant-power amplification. (c) Evolution of the state vector during optimization. This figure illustrates the state vector's behavior as the system iteratively evolves over time. States with higher loss gradually diminish, while states with lower loss persist and are amplified. 

Concretely, we can map optimization problems to the interferometer mesh if relative phase is controlled, than interference between the modes can be used to encode an optimization problem, such that antiferromagnetic coupling is mapped to destructive interference, and ferromagnetic coupling to constructive. This is conceptually similar to a coherent Ising machine encoding but uses the amplitude degrees of freedom rather than phase degrees of freedom to store information. Since amplitude directly corresponds with photon number, Fock state measurements are natural in this setting. A natural way of storing a binary variable for example is in the relative amplitude/number of photons in two modes, such that more photons in one corresponds to $1$ and in the other corresponds to $0$, this is somewhat analogous to the encoding we currently use in the measurement-and-feedback setting.

\begin{figure}[H]
\centering
\includegraphics[width=1\textwidth]{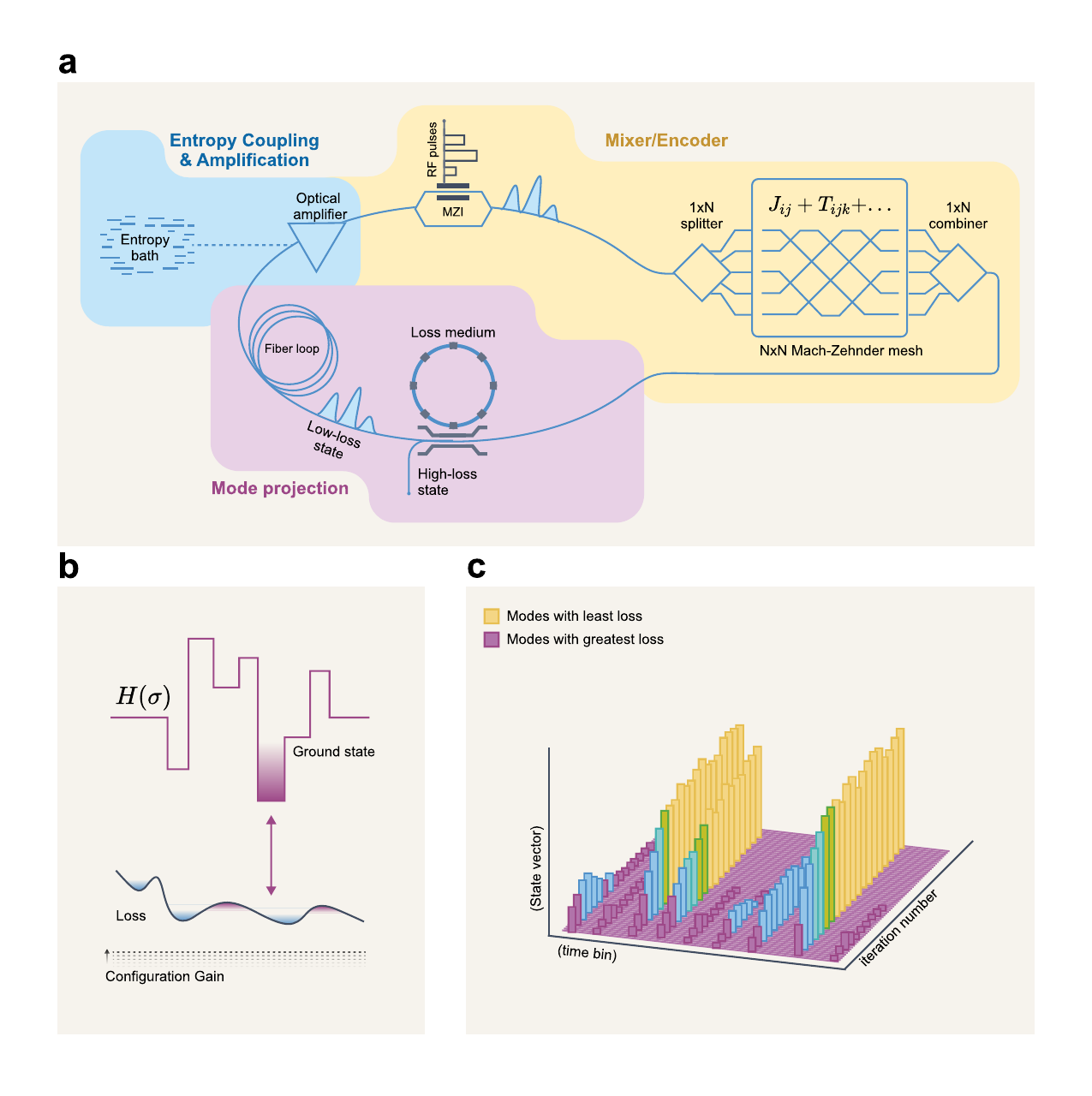}
\textbf{Supplementary Figure 1: An example realization of quantum entropy  computing using time-energy modes and its schematic in a fiber-optical loop.} (b) The principle involves embedding a target Hamiltonian ($\hat{H}$) into a dissipation operator ($-\mathrm{i}\hat{H}$) to drive a quantum state of light toward the ground state while suppressing other states through decay and decoherence. (a) System dynamics starts by amplifying vacuum noise (entropy bath) via optical amplifier, linear terms of Hamiltonian is encoded via electro-optical modulator and coupled through N-by-N Mach Zehnder mesh to realize non-linear interaction of the Hamiltonian. The mixed signals are sent to the loss medium via non-linear waveguide which causes differential loss to each time mode before feeded back into the constant-power amplifier. (c) Evolution of the state vector during optimization. This figure illustrates the state vector's behavior as the system iteratively evolves over time. States with higher loss gradually diminish, while states with lower loss persist and are amplified.  
\label{fig1}
\end{figure}

\section*{Supplementary Note 2: Mapping of Discrete Problems}

To map discrete optimization problems in the objective function of the form described in Eq.~(4) of the main text, we allocate $k$ time bins to a discrete $k$-level variable. For example, in the case of binary variables, e.g., $s_i \in \{-1, 1\}$, we consider two continuous variables to represent a binary variable as $\mathbf{s}_i = (x_i , y_i)^t$, which is a superpositon of the two states $(0, 1)^t$ and $(1, 0)^t$ with non-negative weights $x_i \geq 0$ and $y_i \geq 0$. Here, we consider penalty terms for penalizing results other than the discrete states above. In that case, as an example, the Max-Cut problem on a graph with adjacency matrix $A$, can be formulated as finding state that minimizes the objective function $\sum_{i,j} A_{ij} s_i s_j$, which can be encoded in the following form:
\[
E = \sum_{i,j} A_{ij} \left( x_i x_j + y_i y_j \right) + \lambda \sum_{i}  \left( x_i + y_i - 1 \right)^2.
\]
The constraint term $(x_i + y_i - 1)^2$ tends to force each vector $\mathbf{s}_i = (x_i, y_i)^t$ to be located on the line $x_i + y_i = 1$. This term imposes a regularization for the magnitude of the spin vectors $\mathbf{s}_i$ but does not guarantee the stabilization of individual vectors in the two discrete states of $(0, 1)^t$ and $(1, 0)^t$. The constraint term can be modified to promote such discrete states by adding a product term $x_i y_i$, that is, $(x_i + y_i - 1)^2 \rightarrow (x_i + y_i - 1)^2 + \alpha x_i y_i$. However, in all our simulations the additional term did not improve the results.

The optimization problem can then be cast in the form of Eq.~(4) of the main text, up to quadratic terms, i.e., $E = \mathbf{C}^t \mathbf{v} + \mathbf{v}^t J \mathbf{v}$, where:
\[
\mathbf{v}=\begin{bmatrix}
\begin{matrix}
x_1\\ 
y_1
\end{matrix}\\
\begin{matrix}
x_2\\ 
y_2
\end{matrix} 
\\
\vdots 
\\ 
\begin{matrix}
x_N
\\ 
y_N
\end{matrix}
\end{bmatrix},
%
%
%
J=\begin{bmatrix}
\begin{matrix}
~\lambda~ & ~\lambda~\\ 
~\lambda~ & ~\lambda~
\end{matrix} & \begin{matrix}
a_{12} & 0\\ 
0 & a_{12}
\end{matrix} & \cdots & \begin{matrix}
a_{1N} & 0\\ 
0 & a_{1N}
\end{matrix}\\ 
\begin{matrix}
a_{12} & 0\\ 
0 & a_{12}
\end{matrix} & \begin{matrix}
~\lambda~ & ~\lambda~\\ 
~\lambda~ & ~\lambda~
\end{matrix} & \cdots  & \begin{matrix}
a_{2N} & 0\\ 
0 & a_{2N}
\end{matrix}\\ 
\vdots & \vdots & \ddots  & \vdots \\ 
\begin{matrix}
a_{1N} & 0\\ 
0 & a_{1N}
\end{matrix} & \begin{matrix}
a_{2N} & 0\\ 
0 & a_{2N}
\end{matrix} & \cdots & \begin{matrix}
~\lambda~ & ~\lambda~\\ 
~\lambda~ & ~\lambda~
\end{matrix} 
\end{bmatrix},
%
%
%
\mathbf{C}=\begin{bmatrix}
\begin{matrix}
-2\lambda\\ 
-2\lambda
\end{matrix}\\
\begin{matrix}
-2\lambda\\ 
-2\lambda
\end{matrix} 
\\
\vdots 
\\ 
\begin{matrix}
-2\lambda
\\ 
-2\lambda
\end{matrix}
\end{bmatrix}
\]

This approach can be extended to the standard Potts-k model, $E = \sum_{i,j} A_{ij} \delta \left( \mathbf{s}_i, \mathbf{s}_j \right)$, where, $\mathbf{s}_i$ are in a discrete set with $k$ different elements. In this case, a k-sate vector can be defined as $\mathbf{s}_i = (x_{1i}, x_{2i}, \cdots, x_{ki})^t$, and accordingly an objective function can be defined as follows:
\[
E = \sum_{i,j} A_{ij} \left( x_{i1} x_{j1} + \cdots + x_{ik} x_{jk} \right) + \lambda \sum_{i}  \left( x_{i1} + \cdots + x_{ik} - 1 \right)^2,
\]
which can again be cast in the form of the polynomial objective function described in Eq.~(4) of the main text.

\section*{Supplementary Note 3: Gradient Descent Methodology Used for Comparison with Entropy Computing}
Entropy computing minimizes the energy for a polynomial solution $\mathbf{x}$ subject to following constraints:
\[
\sum_{i} x_i = C, \quad x_i \geq 0
\]

In order to match the constraints used by the entropy computing solver a projection to simplex must be applied at each step of the gradient descent such that:
\[
\Pi_{\Delta_C}(z) = \arg\min_{x} \| x - z \|_2^2, \quad \text{subject to} \quad \sum_{i} x_i = C, \quad x_i \geq 0.
\]

This projection ensures that the solution remains within the same constraints as the entropy computing optimization to allow for comparison. The algorithm for projecting onto the simplex is based on the method introduced by Duchi et al.~\cite{duchi2008efficient}. The method utilizes sorting to efficiently compute the projection in linear time. The projection procedure involves sorting the components of the solution, identifying the threshold for the largest possible values, and ensuring the solution remains within the simplex by enforcing the constraints.

To optimize to optimize the learning rate (\( \eta \)), a Tree-Structured Parzen Estimator (TPE), which is a Bayesian optimization technique widely used for parameter tuning, is employed\cite{watanabe2023treestructuredparzenestimatorunderstanding}. The implementation used in the study was developed by Optuna \cite{DBLP:journals/corr/abs-1907-10902}. The learning rate is optimized with a prior of a log-uniform distribution in the range \([10^{-5}, 10^{-2}]\) over $50$ iterations of sampling to determine the optimal learning rate based on the average energy for 500 randomly initialized solutions. Each gradient descent was run without any stopping conditions for 2000 iterations. After determining the optimal learning rate, this learning rate is then applied to 500 new randomly initialized solutions, which are then used to generate plots and comparisons between the two methods. Further implementation details can be found in the GitHub repository associated with the paper, including versions of packages and the Python version used to generate the data and plots used in the paper at \url{https://github.com/qci-github/eqc-studies/blob/main/eqc-paradigm/eqc_vs_grad}. The same data and code are available at \url{http://doi.org/10.6084/m9.figshare.29949722}.

\section*{Supplementary Note 4: Lack of Gaussian Approximation}

\subsection*{Background}

One issue which has been pointed out in the context of coherent Ising machines is the fact that they are often well approximated by Gaussian states undergoing processes which preserve the Gaussian nature of the states \cite{Clements2017GaussIsing,Tiunov2019simBifurcation,Kosuke2021simBifurcation,Ng2022GaussCIM}. Gaussian states are a special set of optical states which are fully defined through only first and second moments of the quadratures. Through an optical equivalent \cite{Bartlett2002GaussSim} of the famous Gottesmann-Knill theorem \cite{Gottesman1998GKtheorem} of gate-model quantum computing, such systems can be efficiently classically simulated. The analog of these states in gate-model quantum computing are Clifford states and Clifford circuits, which are simulable through stabilizers. 

While Gaussian states and Clifford states are both classically simulable, they are also not fully classical. In fact they are very far from classical in many ways. In the gate model setting, many states such as cluster states are Clifford states and  therefore within this efficiently described space, but are very quantum and play an important role in quantum computing \cite{Nielson2006clusterCompute}. Likewise Gaussian states can be highly entangled.  However, the ease of classical simulations implies that an entirely Clifford or entirely Gaussian system cannot provide a quantum advantage by itself. Anything which the system does can be efficiently simulated on a classical computer.

On the other hand, these easily simulated systems can be used as building blocks toward a quantum advantage, as long as something is added which prevents simulability. For example, combining cluster states with non-Clifford measurements allows for measurement-based quantum computing, which is a universal form of quantum computing \cite{Briegel2009MBQC}. On the optical side, Gaussian Boson sampling \cite{Hamilton2017GBS}, which is believed to be very hard to approximate classically, combines Gaussian states with photon number measurements.

For an optical paradigm of quantum computing to present a true quantum advantage, it should not be completely Gaussian. The usual definition of a coherent Ising machine involves some processes which break the Gaussian nature of the states \cite{Yamamoto2017CIM}. In many cases classical simulations still perform well, this is the paradigm of simulated bifurcation machines \cite{Clements2017GaussIsing,Tiunov2019simBifurcation,Kosuke2021simBifurcation,Ng2022GaussCIM}. Entropy computing involves all of the non-Gaussian mechanisms which are present in the coherent Ising setting, but additionally has number basis measurements.

\subsection*{Coherent Ising Machines}

For the purpose of this section we will consider the all-optical description of coherent Ising machines \cite{Yamamoto2017CIM}. While large systems have not been built to implement this paradigm, it allows us to avoid questions about the classical measurement and feedback. The basic ingredients of an all optical coherent Ising machine consist of lasers, which produce approximate coherent states, a nonlinear component which is able to implement single-mode squeezing, a mesh of beam-splitters and phase shifters which encode the problem, single photon loss which occurs as the states travel through the system, and a final homodyne measurement to read out the Ising states \cite{Yamamoto2017CIM}. 

For squeezed states, loss of single photons does not preserve the Gaussian nature \cite{Ourjoumtsev2006kittens}. Also, while pure squeezing does produce Gaussian states, a more realistic description which includes the possibility of depleting the pump light produces non-Gaussian states \cite{Yamamoto2017CIM,Adesso2014Gaussian}. For these two reasons, an all-optical coherent Ising machine cannot be exactly simulated using Gaussian states and is therefore likely inefficient to exactly describe classically. The immediate question, which is beyond the scope of our current work, is whether the non-Gaussian features which are produced are important to the calculation. In other words, could an approximate Gaussian description of a coherent Ising machine compute as well as (or better than) a real device? 

Computers which are based on simulations of coherent Ising machines are known as simulated bifurcation machines \cite{Kosuke2021simBifurcation}. In general this paradigm has been shown to be similar (and in some cases better\cite{Tiunov2019simBifurcation,Kosuke2021simBifurcation,Ng2022GaussCIM}) in performance to real coherent Ising machines. While this raises some interesting questions about whether coherent Ising machines have potential for a true advantage, it is worth remarking that the only large-scale implementations are based on measurement and feedback.

\subsection*{Entropy Computing}

Now we come to the entropy computing paradigm. As with coherent Ising machines, we consider the all-optical instantiation presented in supplementary Note 1 to avoid questions about destruction of coherent superpositions due to measurement and feedback. While the measurement-and-feedback based version could include non-Gaussian statisitics, the frequent measurement in the computational basis clearly rules out a quantum advantage. In the all optical setting natural loss as seen in a coherent Ising machine is present, and the pumps may be depleted in the non-linear components. In other words, both mechanisms which may free a coherent Ising machine from being effectively described in a Gaussian way are present.

However, there are additional considerations. Since the final measurements in entropy computing are in the number-state basis, these too will break a Gaussian description. Moreover this is very similar to how a Gaussian description is broken in Gaussian Boson sampling\cite{Hamilton2017GBS}. Based on the suspected hardness of simulating Gaussian Boson sampling, even approximately \cite{Grier2022GBScomplexity}, it is reasonable to assume that fully-optical entropy computing will be similarly hard to simulate classically. In fact it is difficult to see how one would even go about constructing an approximate Gaussian model of entropy computing unlike in the coherent Ising settings where there are a clear set of approximations which yield a Gaussian description \cite{Ng2022GaussCIM}. 

Other processes such as Zeno blockade are likely to be able to break the system out of the easy-to-simulate space of Gaussian systems, but given the amount of study on Boson sampling and the relevant analogy with this paradigm, we choose to focus on this route.

\subsubsection*{Role of Bosonic statistics in entropy computing}

We have established that number state measurements provide a novel way to avoid Gaussian simulability within the entropy computing paradigm. The next natural question is whether the Bosonic number statistics which make Gaussian Boson sampling hard play a meaningful role in the ability of the device to optimize. A simple, but we believe still valid, argument here is that richer Bosonic statistics are able to capture correlations which are not easily described classically. These richer statistics are likely to be able to retain more information about the optimization landscape while exploring it. 

It is also worth remarking on what effect photon measurements have on optical systems with squeezing. While difficulty of simulation provides a necessary condition for meaningful beyond-classical computation, it does not necissarily provide a sufficient one. In other words, being difficult to simulate does not automatically imply a performance improvement. However, photon measurement of squeezed states has an interesting effect, it produces states known as Schr{\"o}dinger kitten states \cite{Dakna1997Kitten,Takase2021Kitten,Walschaers2021nonGauss}, which are approximate superpositions of coherent states of opposing phases, effectively such measurements drive superpositions. Therefore while detailed analysis would be needed to conclusively show that this is a source of an advantage, there is a reasonable argument that such measurements could provide the quantum parallelism which squeezing alone cannot (as squeezed states in the absence of measurement can be efficiently described classically).

A state measured at the end would still be fully predicted by the Gaussian description. In other words, even thought the number-basis photon distribution is hard to calculate, the amount of information it holds is still fully determined by the first and second moments of a Gaussian distribution. However, even a few number basis measurements taken earlier in the protocol would render a Gaussian description of the system at later times ineffective. No longer would first and second moments suffice as a full description. Too many measurements of the variables would yield behavior similar to the current measurement-and-feedback approach. However, an intermediate approach where a few number basis measurements are taken early on to induce non-Gaussian behavior but which yield relatively little information within the computational basis are likely to be able to improve performance. Note that similar approaches have been proposed elsewhere, in particular in proposals for variational quantum algorithms for optimization based on Boson sampling \cite{Bradler2021variationalGBS,Goldsmith2024variationalGBS}. A key difference being that the entropy paradigm does not rely on variational techniques and involves inducing non-Gaussian behavior earlier in the protocol as opposed to only right at the end.

One observation is that it is only measurements which (directly or indirectly) give information about the variables which ruin the coherent superposition of values. In fact, some encoding measurements could be devised which would render the system indescribable by Gaussian modes (even approximately) but would not disrupt quantum superpositions of solutions. Consider for example an encoding where a binary variable is encoded into which of two time bins is populated by more photons, if we used a beam splitter to separate off an equal fraction of the light in each bin, and then combined those with a $50:50$ beam splitter the result would tell us nothing about which bin was more occupied but would render the modes in those time bins (and any which they share substantial covariance with) highly non-Gaussian. Effectively the beamsplitter is performing quantum erasure as to which bin the photons originated from, similar to the way in which overlapping beams can restore coherent interference in an induced coherence experiment \cite{Zou1991induced}. The lack of information from the measurement as to which solution is encoded implies that phase coherence between different encoded solutions would be maintained when such a measurement is performed.

\bibliography{sn-bibliography}

\end{document}